\documentstyle[11pt,epsfig]{article}
\begin{document}

\title{Entanglement and Quantum Gate Operations with\\
Spin-Qubits in Quantum Dots
\footnote{To appear in ``Future Trends in Microelectronics: 
The Nano Millenium", eds.
S. Luryi, J. Xu, and A. Zaslavsky, Wiley.}}
\author{John Schliemann$^1$ and Daniel Loss$^2$\\\\
$^1$Department of Physics, The University of Texas, 
Austin, TX 78712\\
$^2$Department of Physics and Astronomy, University of Basel,\\
Klingelbergstrasse 84, 
CH-4056 Basel, Switzerland}

\date{\today}

\maketitle\abstract{We give an elementary introduction to the notion of
quantum entanglement between distinguishable parties 
and review a recent proposal about solid state 
quantum computation with spin-qubits in quantum dots. The indistinguishable 
character of the electrons whose spins
realize the qubits gives rise to further entanglement-like quantum 
correlations. We summarize recent results concerning this type of quantum
correlations of indistinguishable particles.}

\section{Introduction}                         

Quantum entanglement is one of the most intriguing features of quantum 
mechanics \cite{Peres,Steane,Ekert,Werner}. 
In the beginning of modern quantum theory, the notion of 
entanglement was first noted by Einstein, Podolsky, and Rosen \cite{EPR}, 
and by Schr\"odinger \cite{Schroedinger}. While in those days quantum
entanglement and its predicted physical consequences were (at least partially)
considered as an unphysical property of the formalism (a ``paradox''),
the modern perspective on this issue is very different. Nowadays quantum
entanglement is to be seen as an experimentally verfied property of nature
providing a resource for a vast variety of novel phenomena and concepts
such as quantum computation, quantum crytography, or quantum teleportation.

While the basic notion of entanglement in pure quantum states of bipartite
systems (Alice and Bob) is theoretically well understood, fundamental
questions are open concerning entanglement in mixed states (decribed by
a proper density matrix) \cite{Lewenstein,Terhal,Peres2,Horodecki}, 
or entanglement of more than two parties 
\cite{Terhal,Coffman,Acin,Carteret,Duer,Wootters1,OConnor,Thapliyal}.
The most elementary example for entanglement in a pure quantum state is given
by a spin singlet composed from two spin-$\frac{1}{2}$-objects (qubits) owned
by A(lice) and B(ob), respectively,
\begin{equation}
\frac{1}{\sqrt{2}}\left(|\uparrow\rangle_{A}\otimes|\downarrow\rangle_{B}
-|\downarrow\rangle_{A}\otimes|\uparrow\rangle_{B}\right)\,.
\end{equation}
For such a state, the state of the combined system cannot be described by
specifying the state of Alice's and Bob's qubit separately. It is a standard
result of quantum information theory \cite{Peres} that this property does not 
depend on the basis chosen in Alice's or Bob's Hilbert space. 
As we shall see below, the entanglement
of such a quantum state (quantified by an appropriate measure)
is invariant under (independent) changes of basis in both spaces.

Physically measurable consequences of quantum entanglement of the
above kind arise typically (but not exclusively) in terms of two-body
correlations between the subsystems. In this case the effects of
entanglement can typically be cast in terms of so-called Bell
inequalities \cite{Bell} whose violation manifests the presence
entanglement in a given quantum state. Using this formal approach 
the physical existence of quantum entanglement (as opposed to classical
correlations) has unambiguously been verified for the polarization
states of photons by Aspect and
coworkers \cite{Aspect}. Moreover, quantum entanglement is an essential ingredient
of algorithms for quantum computation \cite{Steane,Ekert}, in particular
for Shor's algorithm for decomposing large numbers into their
prime factors \cite{Shor}. This problem is intimately related to public key
cryptograhy systems such as RSA encoding which is widely used in today's
electronic communication.

Among the many proposals for experimental realizations of quantum information
processing solid state systems have the advantage of offering the
perspective to intregrate a large number of quantum gates into a quantum
computer once the single gates and qubits are established. Recently, a
proposal has been put forward invloving qubits formed by the
spins of electrons living on semiconductor quantum dots 
\cite{LossDivincenzo,BLD,QCReview,Schliemann1,Hu}.
In this scenario, the indistinguishable character of the electrons
leads to entanglement-like quantum correlations which require a 
description different from the ususal entanglement between distinguishable
parties (Alice, Bob, ...) in bipartite (or multipartite) systems.
In such a case the proper statistics of the indistinguishable particles has
to be taken into account.

In this article we give an elementary introduction to the notion of
quantum entanglement between distinguishable parties and 
review the aforementioned proposal for quantum computation
with spin-qubits in quantum dots. The indistinguishable 
character of the electrons whose spins
realize the qubits gives rise to further entanglement-like quantum 
correlations. We summarize recent results on the chracterization and
quantification of these quantum correlations which are analogues 
of quantum entanglement between distinguishable parties
\cite{Schliemann1,Schliemann2,Eckert,Li,Paskauskas,Zanardi}.

%%%%%%%%%%%%%%%%%%%%%%%%%%%%%%%%%%%%%%%%%%%%%%%%%%%%%%%%%%%%%%%%%%%%%%%%%%%%%%

\section{Quantum Entanglement between distinguishable parties}

We now give an introduction to basic concepts of characterizing and 
quantifying entanglement between distinguishable parties. We  
concentrate on pure states (i.e. elements of the joint Hilbert space)
of bipartite systems. We then comment only briefly on the case of mixed states
(described by a proper density operator), and entanglement in multipartite
systems. 

One of the most prominent examples of an {\em entangled state} was
already given in the previous section, namely a spin singlet built up
from two qubits. More generally, if Alice and Bob own Hilbert spaces
${\cal H}_{A}$ and ${\cal H}_{B}$ with dimensions $m$ and $n$, respectively,
a state $|\psi\rangle$ is called {\em nonentangled} if it can be
written as a product state, 
\begin{equation}
|\psi\rangle=|\alpha\rangle_{A}\otimes|\beta\rangle_{B}
\label{prodstate}
\end{equation}
with $|\alpha\rangle_{A}\in{\cal H}_{A}$, $|\beta\rangle_{B}\in{\cal H}_{B}$.
Otherwise $|\psi\rangle$ is entangled. The question arises whether a
given state $|\psi\rangle$, expressed in some arbitrary basis of
the joint Hilbert space ${\cal H}={\cal H}_{A}\otimes{\cal H}_{B}$, 
is entangled or not, i.e. whether there are states 
$|\alpha\rangle_{A}$ and $|\beta\rangle_{B}$
fulfilling (\ref{prodstate}). Moreover, one would like to quantify the
entanglement contained in a state vector. 

An important tool to investigate such questions for bipartite systems is the 
biorthogonal Schmidt decomposition \cite{Peres}.
It states that for any state vector $|\psi\rangle\in{\cal H}$
there exist bases of ${\cal H}_{A}$ and ${\cal H}_{B}$ such that
\begin{equation}
|\psi\rangle=\sum_{i=1}^{r}z_{i}
\left(|a_{i}\rangle\otimes|b_{i}\rangle\right)
\label{Schmidt}
\end{equation}
with coeffcients $z_{i}\neq 0$ and the basis states fulfilling
$\langle a_{i}|a_{j}\rangle=\langle b_{i}|b_{j}\rangle=\delta_{ij}$.
Thus, each vector in both bases for ${\cal H}_{A}$ and ${\cal H}_{B}$
enters at most only one product vector in the above expansion.
As a usual convention, the phases of the basis vectors involved in
(\ref{Schmidt}) can be chosen such that all $z_{i}$ are positive.
The expression (\ref{Schmidt}) is an expansion of the state $|\psi\rangle$
into a basis of product vectors $|a\rangle\otimes|b\rangle$ with a minimum 
number $r$ of nonzero terms. This number ranges from one to
$\min\{m,n\}$ and is called the {\em Schmidt rank} of $|\psi\rangle$.

With respect to arbitrary bases in  ${\cal H}_{A}$ and ${\cal H}_{B}$ a
given state vector reads
\begin{equation}
|\psi\rangle=\sum_{a,b}M_{ab}|a\rangle\otimes|b\rangle
\end{equation}
with an $m\times n$ coefficient matrix $M$. Under unitary transformations
$U_{A}$ and $U_{B}$ in ${\cal H}_{A}$ and ${\cal H}_{B}$, respectively, $M$
transforms as
\begin{equation}
M\mapsto M'=U_{A}MU_{B}^{T}
\end{equation}
with $U_{B}^{T}$ being the transpose of $U_{B}$. The fact that there are always
bases in ${\cal H}_{A}$ and ${\cal H}_{B}$ providing a biorthogonal
Schmidt decomposition of $|\psi\rangle$ is equivalent to stating that
there are matrices $U_{A}$ and $U_{B}$ such that the resulting matrix
$M'$ consists of a diagonal block with only nonnegative entries while the
rest of the matrix contains only zeroes. For the case of equal dimensions
of Alice's and Bob's space, $m=n$, this is also a well-known theorem of
matrix algebra \cite{Mehta}.

Obviously, $|\psi\rangle$ is nonentangled, i.e. a simple product state,
if and only if its Schmidt rank is one. More generally, the Schmidt rank of
a pure state can be viewed as a rough characterization for its entanglement. 
However, since the Schmidt rank is by construction a discrete quantity 
it does not provide a proper quantification of entanglement. Therefore
finer entanglement measures are desirable. 
For the case of two distinguishable parties, a useful measure of
entanglement is the von Neumann-entropy of partial density matrices
constructed from the pure-state density matrix $\rho=|\psi\rangle\langle\psi|$
\cite{BBP+:96} :
\begin{equation}
E(|\psi\rangle)=-{\rm tr}_{A}\left(\rho_{A}\log_{2}\rho_{A}\right)
=-{\rm tr}_{B}\left(\rho_{B}\log_{2}\rho_{B}\right)\,,
\label{vNeuEnt}
\end{equation}
where the partial density matrices are obtained by tracing out one of the
subsystems, $\rho_{A/B}={\rm tr}_{B/A}\rho$. With the help of the 
biorthogonal Schmidt decomposition of $|\psi\rangle$ one shows that
both partial density matrices have the same spectrum and therefore the 
same entropy, as stated in Eq.~(\ref{vNeuEnt}). In particular, the
Schmidt rank of $\psi\rangle$ equals the algebraic rank of the partial
density matrices. $|\psi\rangle$ is nonentangled if and only if the
partial density matrices of the pure state $\rho=|\psi\rangle\langle\psi|$ 
are also pure states, and $|\psi\rangle$ is maximally entangled if its
partial density matrice are ``maximally mixed'', i.e. if they have only
one non-zero eigenvalue with a multiplicity of $\min\{m,n\}$.

It is important to observe that the entanglement
measure (\ref{vNeuEnt}) of a given state $|\psi\rangle$
does not depend on the bases used in Alice's and Bob's Hilbert space
to express this state. 
This is because the trace operations in the definition of
$E(|\psi\rangle)$ are invariant under an eventual change
of bases (performed, in general, independently in both spaces).
Therefore, entanglement in bipartite systems is a basis-independent
quantity.

Thus, the problem of characterizing and quantifying 
quantum entanglement for pure states in bipartite systems can been
seen as completely solved. Unfortunately, the situation is much less clear
for mixed states \cite{Lewenstein,Terhal,Peres2,Horodecki}, 
and for multipartite entanglement. The main obstacle in the latter issue
is the fact that the biorthogonal Schmidt 
decomposition in bipartite systems does not have a true analogue
in the multipartite case. For details we refer the reader to the  
reserach literature; a nonexhaustive colllection of recent papers includes
\cite{Terhal,Coffman,Acin,Carteret,Duer,Wootters1,OConnor,Thapliyal}.

%%%%%%%%%%%%%%%%%%%%%%%%%%%%%%%%%%%%%%%%%%%%%%%%%%%%%%%%%%%%%%%%%%%%%%%%%%%%%%

\section{Quantum computing with electron spins in quantum dots}

We will now illustrate the phenomenon of quantum entanglement on the
example of a specific (possible) realization of a quantum information
processing system \cite{LossDivincenzo}. The proposal discussed below
deals with qubits realized by the spins of electrons residing on
semiconductor quantum dots. As we shall see in this and the following
section, the indistinguishable character of the electrons gives rise
to quantum correlations which are beyond entanglement between distinguishable
parties.

\begin{figure}[t]
\centerline{\includegraphics[width=12cm]{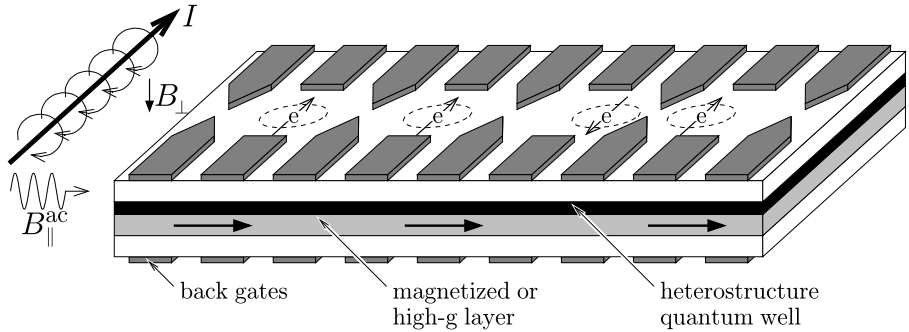}}
\caption{\label{figArray}
Quantum dot array, controlled by electrical gating.
The electrodes (dark gray) define quantum dots (circles) by confining electrons
The spin 1/2 ground state (arrow) of the dot represents the qubit.
These electrons can be moved by electrical gating into the magnetized or 
high-$g$ layer, producing locally different Zeeman splittings.
Alternatively, magnetic field gradients can be applied, as e.g. produced
by a current wire (indicated on the left of the dot-array).
Then, since every dot-spin is subjected to a different Zeeman
splitting, the spins can be addressed individually, e.g. 
through ESR pulses of an additional in-plane magnetic ac field
with the corresponding 
Larmor frequency $\omega_{\rm L}=g\mu_{B}B_\perp/\hbar$.
Such mechanisms can be used for single-spin rotations and the
initialization step.
The exchange coupling between the quantum dots can be controlled by
lowering the tunnel barrier between the dots.
In this figure, the two rightmost dots are drawn schematically as 
tunnel-coupled. Such an exchange mechanism can be used for the XOR
gate operation involving two nearest neighbor qubits. The XOR operation
between distant qubits is achieved by swapping (via exchange)  the qubits
first to
a  nearest neighbor position. 
The read-out of the spin state can be achieved via 
spin-dependent tunneling and SET devices \cite{LossDivincenzo}, 
or via a transport current passing the dot \cite{EL}.
Note that all spin operations, single and two
spin operations, and spin read-out, are controlled electrically via the
charge of the electron and not via the magnetic moment of the spin. Thus, no
control of local magnetic fields is required, and the spin is only used for
storing the information.
This spin-to-charge conversion is based on the Pauli principle and Coulomb
interaction and allows for very fast switching times (typically picoseconds). 
A further advantage of this scheme is its scalability into an array of arbitrary
size. }
\end{figure}

An array of coupled quantum dots, see Fig.~\ref{figArray}, each dot containing
a top most spin 1/2, was found to be a promising candidate for a scalable 
quantum computer~\cite{LossDivincenzo} where the quantum bit (qubit) is 
defined by the spin $1/2$  on the dot. Quantum algorithms can then be 
implemented using 
local single-spin rotations and the exchange coupling between nearby spins,  
see Fig.~\ref{figArray}. This proposal is supported by experiments where, 
e.g., Coulomb blockade effects,~\cite{waugh}
tunneling between neighboring dots,~\cite{Kouwenhoven,waugh}
and magnetization~\cite{oosterkamp} have been observed as well as the
formation of a delocalized single-particle state in coupled dots
\cite{blick}. For a detailed review of quantum computing with electron spins 
in quantum dots see Ref.~\cite{QCReview}.

The charge of the electron can further be used to transport a spin-qubit along
conducting wires \cite{BLS}.
This allows one to use  spin-entangled electrons as
Einstein-Podolsky-Rosen (EPR) \cite{EPR} pairs, which
can be created
(e.g.\ in coupled quantum dots or near a superconductor-normal interface),
transported, and detected in transport and noise measurements 
\cite{BLS,RSL,LS}. Such EPR pairs represent the fundamental resources for quantum
communication
\cite{Bennett00}.
The electron spin is a natural candidate for a qubit since its spin state in 
a given direction, $|\uparrow\rangle$ or $|\downarrow\rangle$, 
can be identified with the classical 
bits $|0\rangle$ and $|1\rangle$, while an arbitrary superposition 
$\alpha\|\uparrow\rangle+\beta\|\downarrow\rangle$ defines a qubit.
In principle, any quantum two-level system can be used to define a qubit.
However, one must be able to control coherent superpositions of
the basis states of the quantum computer,
i.e. no transition from quantum to classical behavior should occur.
Thus, the coupling of the environment to the qubit should be small,
resulting in a sufficiently large decoherence time $T_2$
(the time over which the phase of a superposition of $|0\rangle$
and  $|1\rangle$ is well-defined).
Assuming weak spin-orbit effects,
the spin decoherence time $T_{2}$ can be completely different
from the charge decoherence time (a few nanoseconds),
and in fact it is known~\cite{Kikkawa} that $T_{2}$ can be orders of 
magnitude longer than nanoseconds. 
Time-resolved optical measurements were used to determine $T_2^{*}$,
the decoherence time of an ensemble of spins, with $T_2^{*}$
exceeding 100 ns in bulk GaAs \cite{Kikkawa}.
More recently, the single spin relaxation time $T_1$
(generally $T_1 \geq T_2$) of a single quantum dot attached to leads
was measured via transport to be longer than a few $\mu$s \cite{Fujisawa},
consistent with calculations \cite{KhaetskiiNazarov}.

Let us now consider a system of two laterally tunnel-coupled dots
having one electron each. Using an appropriate model 
\cite{BLD,Schliemann1}
theoretical calculations have demonstrated the possibility of performing
two-qubit quantum gate operations in such a system by varying the 
tunnel barrier between the dots. An important point to observe here 
is the fact that the electrons whose spins realize the qubits are
indistinguishable particles \cite{Schliemann1}. Differently from  
the ususal scenario of distinguishable parties (Alice, Bob, ...)
the proper quantum statistics has to be taken into account when a finite
tunneling between the dots is inferred \cite{Schliemann1,Schliemann2,Eckert}. 

In the following section we give an elementary introduction to
the theory of ``entanglement-like'' quantum correlations in systems of
indistinguishable particles. We concentrate on the fermionic case and
illustrate our findings on the above example of coupled quantum dots.

%%%%%%%%%%%%%%%%%%%%%%%%%%%%%%%%%%%%%%%%%%%%%%%%%%%%%%%%%%%%%%%%%%%%%%%%%%%%%%

\section{Quantum Correlations between indistinguishable particles}

For indistinguishable particles  a pure
quantum state must be formulated in terms of Slater determinants or
Slater permanents for fermions and bosons, respectively.
Generically, a Slater determinant contains correlations due to the exchange 
statistics of the indistinguishable fermions. As the simplest possible
example consider a wavefunction of two (spinless) fermions,
\begin{equation}
\Psi(\vec r_{1},\vec r_{2})=\frac{1}{\sqrt{2}}
\left[\phi(\vec r_{1})\chi(\vec r_{2})-\phi(\vec r_{2})\chi(\vec r_{1})
\right]
\label{fermwv}
\end{equation}
with two orthonormalized single-particle wavefunctions 
$\phi(\vec r)$, $\chi(\vec r)$. Operator matrix elements between such 
single Slater
determinants contain terms due to the antisymmetrization of coordinates
(``exchange contributions'' in the language of Hartree-Fock theory).
However, if the moduli of $\phi(\vec r)$, $\chi(\vec r)$ have only
vanishingly small overlap, these exchange correlations will also tend
to zero for any physically meaningful operator. This situation is generically
realized if the supports of the single-particle wavefunctions are essentially
centered around locations being sufficiently apart from each other, or the
particles are separated by a sufficiently large energy barrier. 
In this case the
antisymmetrization present in Eq.~(\ref{fermwv}) has no physical effect.

Such observations clearly 
justify the treatment of indistinguishable particles separated by macroscopic
distances as effectively distinguishable objects. So far, research in 
Quantum Information Theory has concentrated on this case, where the exchange
statistics of particles forming quantum registers could be neglected, or was 
not specified at all. 

The situation is different if the particles constituting, say, qubits are
close together and possibly coupled in some computational process.
This the case for all proposals of quantum information
processing based on quantum dots technology 
\cite{LossDivincenzo,BLD,QCReview,Schliemann1,Hu}. Here qubits are
realized by the spins of electrons living in a system of quantum dots.
The electrons have the possibility of tunneling eventually from one dot to the
other with a probability which can be modified by varying external parameters
such as gate voltages and magnetic field. In such a situation the fermionic
statistics of electrons and the associated Pauli principle are clearly essential. 

Additional correlations in many-fermion-systems
arise if more than one Slater determinant  is involved, i.e. if
there is no single-particle basis such that a
given state of $N$ indistinguishable fermions can be
represented  as an elementary  Slater determinant (i.e. fully antisymmetric
combination of $N$ orthogonal single-particle states).
These correlations are  the analog of quantum entanglement in 
separated systems and are essential for
quantum information processing in non-separated systems.

As an example consider a ``swap'' process exchanging the spin states of 
electrons on coupled quantum dots by gating the tunneling amplitude between 
them \cite{BLD,Schliemann1}. 
Before the gate is turned on, the two electrons in
the neighboring quantum dots are in a state represented by 
a simple Slater determinant, and can be regarded as distinguishable since they
are separated by a large energy barrier. When the barrier is lowered, 
more complex correlations between the electrons due to the dynamics arise. 
Interestingly, as shown in Refs. \cite{BLD,Schliemann1}, 
during such a process the system must necessarily enter a highly
correlated  state that cannot be represented by a single Slater determinant. 
The final state of the gate operation, however, 
is, similarly as the initial 
one, essentially given by a single Slater determinant. 
Moreover, by adjusting the gating time
appropriately one can also perform a ``square root of a swap'' which turns
a single  Slater determinant into a ``maximally'' correlated state
in much the same way \cite{Schliemann1}. Illustrative details of 
these processes will be given below. 
In the end of such a process the electrons
can again be viewed as effectively distinguishable, but are in 
a maximally entangled state in the usual sense of 
distinguishable separated particles. In this sense the highly correlated
intermediate state can be viewed as a resource for the production of
entangled states.

In the following we give an elementary introduction to recent results
in the theory of quantum correlations in systems of indistinguishable
particles \cite{Schliemann1,Schliemann2,Eckert,Li,Paskauskas}. 
These correlations are analogues of entanglement between
distinguishable parties.
However, to avoid confusion with the existing literature and in accordance
with Refs. \cite{Schliemann2,Eckert,Paskauskas},
we shall reserve in the following the term ``entanglement'' for separated
systems and characterize the analogous quantum correlation phenomenon in 
nonseparated systems in terms of the  
Slater rank and the correlation measure 
to be defined below. 

For the purposes of this article we shall concentrate on elementary 
results for the case of pure states of 
two identical fermions. Results for mixed states and more than two fermions
can be found in \cite{Schliemann2,Eckert}. Results for the case 
of identical bosons can be found in \cite{Li,Paskauskas,Eckert}

We consider the case of two identical fermions sharing an
$n$-dimensional single-particle space ${\cal H}_{n}$ resulting in a 
total Hilbert space ${\cal A}({\cal H}_{n}\otimes{\cal H}_{n})$ with
${\cal A}$ denoting the antisymmetrization operator. A general state vector
can be written as
\begin{equation}
|w\rangle=\sum_{a,b=1}^{n}w_{ab}f^{+}_{a}f^{+}_{b}|0\rangle
\label{defstate}
\end{equation}
with fermionic creation operators $f^{+}_{a}$ acting on the vacuum
$|0\rangle$. The antisymmetric coefficient
matrix $w_{ab}$ fulfills the normalization condition 
\begin{equation}
{\rm tr}\left(\bar w w\right)=-\frac{1}{2}\,,
\end{equation}
where the bar stands for complex conjugation. Under a unitary transformation 
of the single-particle space,
\begin{equation}
f^{+}_{a}\mapsto{\cal U}f^{+}_{a}{\cal U}^{+}=U_{ba}f^{+}_{b}\,,
\end{equation}
$w$ transforms as 
\begin{equation}
w\mapsto UwU^{T}\,,
\end{equation}
where $U^{T}$ is the transpose (not the adjoint) of $U$. For any complex
antisymmetric matrix $n\times n$ matrix $w$ there is a unitary transformation
$U$ such that $w'=UwU^{T}$ has nonzero entries only in $2\times 2$ 
blocks along the diagonal \cite{Schliemann2,Mehta}. That is,
\begin{equation}
w'={\rm diag}\left[Z_{1},\dots,Z_{r},Z_{0}\right]
\quad{\rm with}\quad Z_i=\left[
\begin{array}{cc}
0 & z_i \\
-z_i & 0
\end{array}
\right]\,,
\label{z}
\end{equation}
$z_{i}\neq 0$ for $i\in\{1,\dots,r\}$, and $Z_{0}$ being the 
$(n-2r)\times(n-2r)$ null matrix.
Each $2\times 2$ block $Z_{i}$ corresponds to an elementary Slater 
determinant in
the state $|w'\rangle$. Such elementary Slater determinants are the analogues
of product states in systems consisting of distinguishable parties.
Thus, when expressed in such a basis, the state $|w\rangle$ 
is s sum of elementary Slater determinants 
where each single-particle basis state enters not more
than one term. This property is analogous to the biorthogonality of the 
Schmidt decomposition discussed above.
The matrix (\ref{z}) represents an expansion of $|w\rangle$ into a basis of
elementary Slater determinants with a minimum number $r$ of non-vanishing
terms. This number is analogous to the Schmidt rank for the distinguishable
case. Therefore we shall call it the {\em (fermionic) Slater rank} of 
$|w\rangle$ \cite{Schliemann2}, and an expansion of the above form a
{\em Slater decomposition} of $|w\rangle$. 

We now turn to the case of two fermions in a four-dimensional single-particle
space. This case is realized in a system of two coupled quantum dots
hosting in total two electrons which are restricted to the lowest 
orbital state on each dot. In such a system, a simple correlation
measure can be defined as follows \cite{Schliemann1,Schliemann2}: 
For a given state (\ref{defstate}) with a coefficient matrix
$\omega_{ab}$ one defines a dual state $|\tilde\omega_{ab}\rangle$
characterized by the dual matrix
\begin{equation}
\tilde w_{ab}=\frac{1}{2}\sum_{c,d=1}^{4}\varepsilon^{abcd}\bar w_{cd}\,,
\label{defdual}
\end{equation}
with $\epsilon^{abcd}$ being the usual totally antisymmetric unit tensor.
Then the {\em correlation measure}
$\eta(|w\rangle)$ can be defined as
\begin{equation}
\eta(|w\rangle)=\left|\langle\tilde w|w\rangle\right|
=\left|\sum_{a,b,c,d=1}^{4}
\varepsilon^{abcd}w_{ab}w_{cd}\right|
=\left|8\left(w_{12}w_{34}+w_{13}w_{42}+w_{14}w_{23}\right)\right|\,.
\end{equation}
Obviously, $\eta(|w\rangle)$ ranges from zero to one. Importantly it vanishes
if and only if the state $|w\rangle$ has the fermionic Slater rank one, i.e.
$\eta(|w\rangle)$
is an elementary Slater determinant. This statement was proved first in
Ref.~\cite{Schliemann1}; an alternative proof can be given using the 
Slater decomposition of $|w\rangle$ and observing that
\begin{equation}
\det w=\left(\frac{1}{8}\langle\tilde w|w\rangle\right)^{2}\,.
\label{det}
\end{equation}
The quantity $\eta(|w\rangle)$ measure quantum correlation contained
in the two-fermion state $|w\rangle$ which are beyond simple
antisymmtrization effects. This correlation measure in under many aspects
analogous to the entanglement measure ``concurrence'' used in systems
of two distinguishable qubits \cite{Wootters2}. These analogies are
discussed in detail in \cite{Eckert} including also the case of
indistinguishable bosons. An important difference between just 
two qubits, i.e. two distinguishable two-level systems,
and the present case of two electrons in a two-dot system is
that in latter system both electrons can eventually occupy the same dot
while the other is empty. Therefore the total Hilbert space is larger
than in the two-qubit system, and a generalized correlation measure
becomes necessary. Furthermore, similar as in the two-qubit case, the
correlation measure $\eta$ defined here for pure states of two fermions 
has a natural extension to mixed fermionic and bosonic states
\cite{Schliemann2,Eckert}.

The expansion of the form (\ref{z}) for a two-fermion system
has an analogue in two-boson systems which was presented very recently in
Refs. \cite{Li,Paskauskas}. 
Moreover the fermionic analogue (\ref{z}) of the biorthogonal
Schmidt decomposition of bipartite systems was also used earlier in
studies of electron correlations in Rydberg atoms \cite{atomphys} 

We note that the aforementioned double occupancies have temporarily given 
raise to some controversy about the principle suitability of such systems as
quantum gates; these concerns were eliminated in a recent theoretical study,
see \cite{Schliemann1} and references therein.

Let us now have a closer look at a specific quantum gate operation, 
namely a swap process oulined already before. This operation 
interchanges the contents of the qubits on two dots, e.g.,
\begin{equation}
|\uparrow\downarrow\rangle\mapsto|\downarrow\uparrow\rangle\,,
\end{equation}
where obvious notation has been used for the spin state on each dot. 
As we shall see below, the ``square root'' of such a swap operation
provides an efficient way to generate entangled states. Moreover,
the ``square root of a swap'' can be combined with further single-qubit
operations to an exclusive-OR (XOR, or controlled-NOT) gate, which has
been shown to be sufficient for the implementation of any quantum
algorithm \cite{DiVincenzo}.

\begin{figure}
\centerline{\includegraphics[width=8cm]{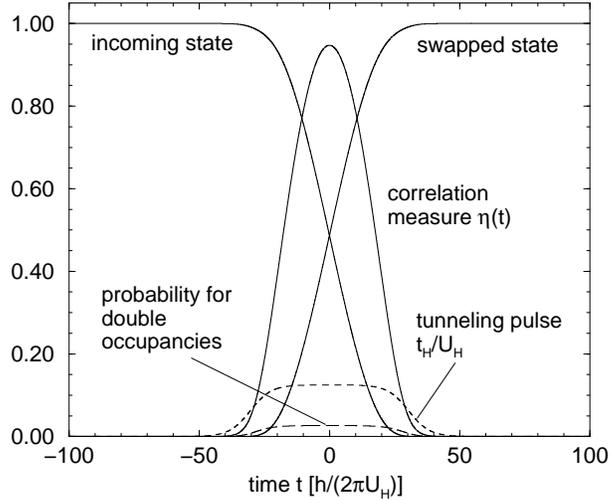}}
\caption{A swap process as a function of time. The 
tunneling amplitude $t_{H}(t)$ is plotted in units of the effective
repulsion $U_{H}$ between elctrons on the same dot.
The square amplitude of the incoming state $|\uparrow\downarrow\rangle$
and the outgoing state  $|\downarrow\uparrow\rangle$ are
shown as thick lines. The probability to find both electrons on the same dot
is necessarily finite $\em during$ the swap process but exponentially
suppressed after it.
The measure of entanglement $\eta(t)$ is also shown.
\label{fig2}}
\end{figure}

Both the initial and the final state in the above example of a quantum
gate operation are single Slater determinants. In the
beginning and the end of the operation the tunneling amplitude between the
dots is exponentially small. The swap process is performed by
temporarily gating the tunneling with a pulse-shaped time dependence
as shown in Fig.\ref{fig2}. In the presence of a finite tunneling 
amplitude, i.e. during the swap operation, a finite probability for
both electrons being on the same dot necessarily occurs. However, this
double occupancy probability can be suppressed very efficiently 
{\em in the final swapped state} provided that the dynamics of the
system is sufficiently close to its adiabatic limit. In fact, as shown
in Ref.~\cite{Schliemann1} this quasi-adiabatic regime is remarkably large.
As a result, a clean swap process can be performed even if the tunneling
pulse is switched on and off on a time scale close to the
natural time scale of the problem given by $\hbar/U_{H}$ where $U_{H}$ is
an effective repulsion between electrons on the same dot. In the middle of the
swap process the system is in an highly correlated quantum state with the
correlation measure $\eta$ being close to its maximum. 

\begin{figure}
\centerline{\includegraphics[width=8cm]{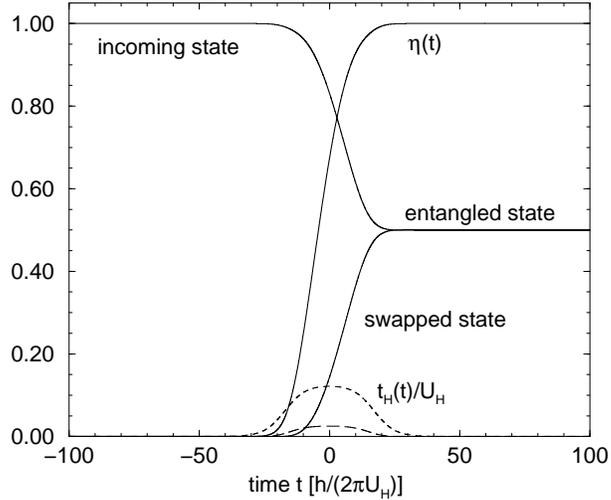}}
\caption{A square root of a swap, which is obtained from the situation of
figure \protect{\ref{fig2}} by halfing the pulse duration $T$. The probability
of double occupancies is again strongly suppressed after the tunneling pulse.
The resulting state is a fully correlated complex linear combination of 
the incoming state $|\uparrow\downarrow\rangle$
and the outgoing state  $|\downarrow\uparrow\rangle$.
of the full swap. The quantum mechanical weigths 
of the latter states are plotted as thick solid lines.
\label{fig3}}
\end{figure}

Next let us look at the ``square root of a swap'', which is obtained from the 
situation of
Fig.~\ref{fig2} by halfing the pulse duration $T$. The probability
of double occupancies is again strongly suppressed after the tunneling pulse.
As shown in Fig.~\ref{fig3},
the resulting state is a maximally correalted ($\eta=1$)
complex linear combination of 
the incoming state $|\uparrow\downarrow\rangle$
and the outgoing state  $|\downarrow\uparrow\rangle$
of the full swap with both states having the same weight.
The quantum mechanical weigths 
of the latter states are plotted as thick solid lines. 
{\em After} the tunneling amplitude is switched off again to exponentially 
small values,  both dots carry one 
electron each. As explained above, in this situation, due to the high
tunneling barrier between the dots, the two electrons can be
considered as effectively distinguishable. In this sense the resulting
state in Fig.~\ref{fig3} can be seen as a ususal entangled state
for distinguishable parties. However, {\em during} the gate operation
such a view is not possible, since there is necessarily a finite amplitude
for doubly occupied dots. Since the amplitude of such spin singlet states
contributes to the
correlation measure $\eta(t)$, the intermediate state during the gate
operations shown in Figs.~\ref{fig2},\ref{fig3} can, loosely speaking,
be interpreted to contain spin as well as orbital entanglement.
In the end of the square root of the swap, however, the correlations
are purely due to the spin degree of freedom.
As a result, the double dot two-qubit system is also an efficient entangler.

%%%%%%%%%%%%%%%%%%%%%%%%%%%%%%%%%%%%%%%%%%%%%%%%%%%%%%%%%%%%%%%%%%%%%%%%%%%%%%

\section{Summary}

We have given an elementary introduction to the notion of quantum
entanglement between distinguishable parties. Entanglement phenomena
can be illustrated on the example of the recently proposed
spin-qubits in quantum dots \cite{LossDivincenzo}. As long as each dots
carries one (valence) electron only with high barriers to the neighboring dots,
the particles constituting the qubits can be seen as effectively 
distinguishable, and the ususal concept and theory of quantum
entanglement applies.
However, two-qubit quantum gate operations in such systems  
are performed by temporarily lowering the tunneling barriers. In such a
situation, the indistinguishable character and proper statistics of
the electrons have to be taken into account. 
Therefore, the question arises
how to describe ``entanglement'' (or, more precisely, quantum
correlations analogous to entanglement between distinguishable parties)
in systems of indistinguishable particles. In this article we have 
provided a simple introduction to this kind of questions and reported
on some elementary results. Interesting questions for further research
include experimental manifestations of entanglement-like quantum
correlations between fermions using the full antisymmetrized
(or, in the case of bosons, symmetrized) Fock space. In particular,
possible generalizations of Bell inequalities to the case of indistinguishable
particles might complement the approach outlined in this article and
suggest experimental studies and applications.\\

We thank D. Bru{\ss}, G. Burkard, J.~I. Cirac, K. Eckert, M. Ku{\'s}, 
M. Lewenstein, A.~H. MacDonald, W.~K. Wootters, L. You, 
and P. Zanardi for useful discussions.
This work  was supported  by the Deutsche Forschungsgemeinschaft,
the Robert A. Welch Foundation,  the US  (DMR0115927) and Swiss NSF, 
DARPA, and ARO.

\end{document}